# Science et politique aux États-Unis[*]

James D. Wells
*Physics Department & Ford School of Public Policy*
*University of Michigan, Ann Arbor*

C'est un grand honneur de pouvoir prendre la parole à l'occasion des 150 ans de la Société Française de Physique. Au moment de sa fondation, la révolution scientifique de la mécanique de Newton avait presque 100 ans, mais les graines de sa destruction avaient déjà été plantées par l'analyse méticuleuse de Le Verrier de l'orbite de Mercure, publiée en 1859. D'autres domaines de la physique classique étaient également en pleine floraison. L'électricité et le magnétisme étaient enfin unifiés en une grande théorie de Maxwell après de gros travaux préparatoires effectués par les physiciens français Coulomb, Ampère, Biot, et Fresnel, entre autres.

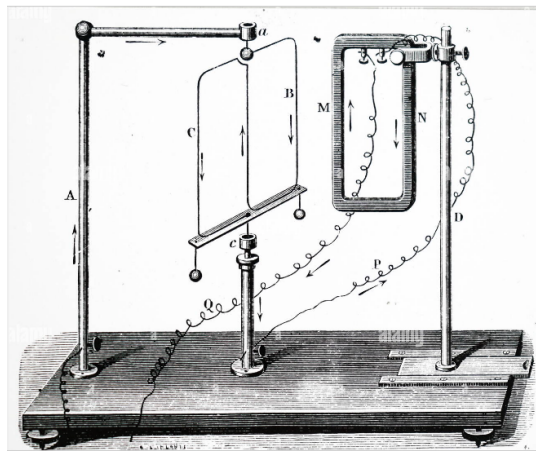

L'expérience d'Ampère

---

[*] Cette présentation a été faite lors du Congrès Général des 150 ans de la Société Française de Physique, Paris, 6 juillet 2023.

A cette époque, donc, les découvertes abstraites du passé se transformaient rapidement en technologies du présent et du futur. Le grand romancier Jules Verne avait bien capturé l'esprit de cette époque, capitalisant sur le lien nouvellement apprécié entre les découvertes scientifiques fondamentales et les avantages pratiques futurs. C'est le premier point majeur que je souhaite aborder dans cet exposé : le soutien politique à la recherche scientifique fondamentale aux États-Unis repose de plus en plus sur la perception de sa valeur directe pour l'ensemble de la société. Elle fait l'objet d'un examen sans précédent compte tenu des récentes pressions culturelles et financières.

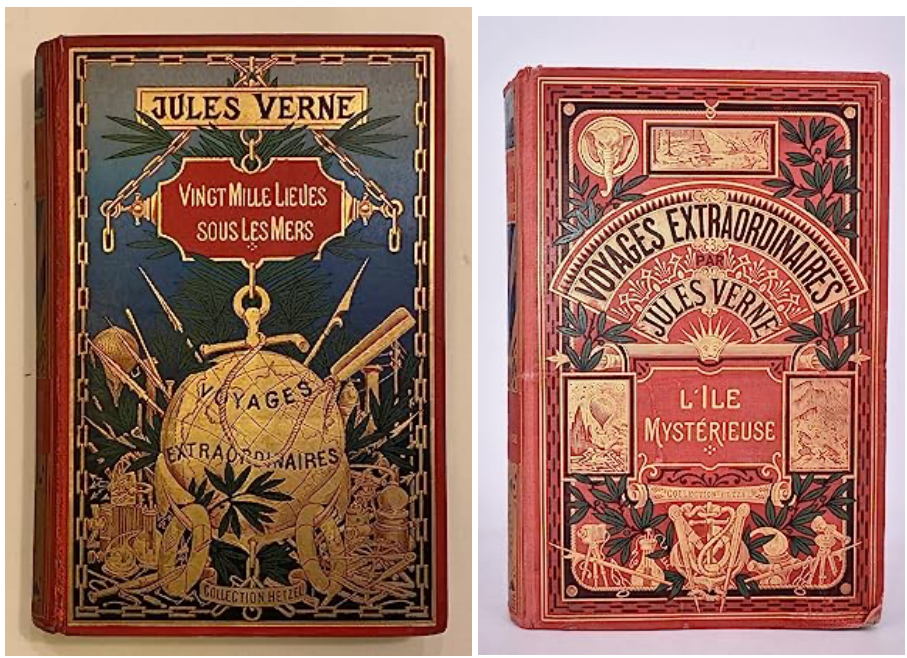

*La vision changeante de la recherche scientifique*

Les États-Unis sont en train de s'éloigner de la recherche purement motivée par la curiosité pour se concentrer de plus en plus sur la recherche axée sur les applications. Récemment, le Congrès a demandé à la National Science Foundation, ou NSF, de créer une nouvelle direction appelée « Technology, Innovation, and Partnerships ». Ils lui ont confié « la mission essentielle de faire progresser la compétitivité des États-Unis grâce à des investissements qui accélèrent le développement de technologies clés et répondent aux défis sociétaux et économiques urgents ». Cette nouvelle direction est controversée et vise à faire comprendre que la curiosité pure des lois fondamentales de la nature n'est pas l'objectif principal du soutien gouvernemental. Au contraire, l'objectif principal est le bénéfice direct pour la société.

Ces développements récents peuvent être comparés au rapport visionnaire de Vannevar Bush de 1945 – un rapport qui avait été commandé par le président Roosevelt, et dans lequel Bush écrivît :
> « Le progrès scientifique sur un large front résulte du libre jeu d'intellects libres, travaillant sur des sujets de leur choix, de la manière dictée par leur curiosité pour l'exploration de l'inconnu. La liberté d'enquête doit être préservée dans les tous les plans gouvernementaux de soutien aux sciences... »

Bush y soulignait que « les nouveaux produits et processus ne naissent pas à maturité. Ils sont fondés sur de nouveaux principes et de nouvelles conceptions qui, à leur tour, résultent de la recherche scientifique fondamentale ».

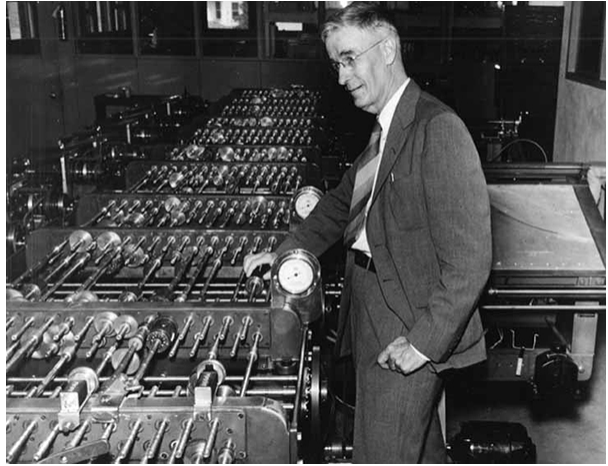

Vannevar Bush

Bush reconnaissait que l'Europe était le leader de la recherche scientifique fondamentale, sur laquelle nous, Américains, avions compté, alors que l'Europe gisait blessée après une guerre catastrophique sur son continent. Il faisait remarquer alors que « nous ne pouvons plus dépendre de l'Europe en tant que source majeure de ce capital scientifique [de la recherche scientifique de base]. »

Le Congrès américain réagi rapidement et la NSF fut créée en 1950, avec, comme charte et objectifs, la vision du rapport Vannevar Bush. La science fondamentale avait ainsi obtenu un soutien solide. Et encore, récemment, c'est la NSF qui a financé la détection des ondes de gravité. Ceci avait été une décision à haut risque, mais elle a été extrêmement payante au sein de la communauté scientifique. Elle a ouvert une toute nouvelle fenêtre sur l'histoire et la structure de notre univers. Cependant, à présent, le Congrès dit que de telles découvertes révolutionnaires ne suffisent plus.

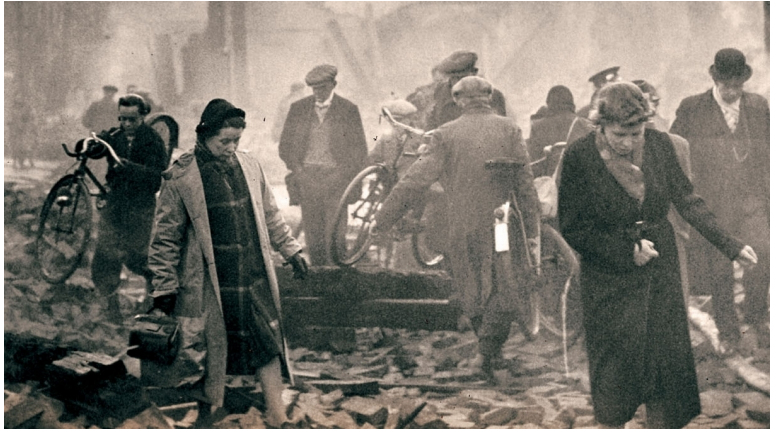

Une guerre catastrophique

L'environnement pourrait bientôt être pire que même en 2017 pour la science fondamentale, lorsque le président Trump avait déjà tenté d'instituer des coupes budgétaires massives dans les portefeuilles de recherche des projets de science fondamentale. Le Congrès, qui a la seule autorité en matière de crédits, a finalement ignoré sa tentative à ce moment-là. Cependant, des baisses soudaines et substantielles du financement de la science, ou des changements soudains de priorité, pourraient avoir lieu après la prochaine élection. Cela dépendra des candidats qui remporteront la présidence et du nouveau Congrès.

Les plus hauts niveaux de gouvernement examinent de plus en plus attentivement la manière dont les projets scientifiques à grande échelle aident directement leurs partisans.

*Les tensions alimentent le changement*

Les origines de ce changement proviennent de plusieurs tensions. Premièrement, le pays est profondément divisé politiquement. Les démocrates et les républicains semblent parfois provenir de planètes différentes. Cela conduit à un effet répulsif, où, si un côté est en faveur d'une politique, l'autre côté ne peut pas l'être aussi. Les demandes budgétaires de l'administration Trump pour la science montrent que le parti républicain, traditionnellement fervent partisan de la science, pourrait abandonner le consensus bipartite sur le soutien à la science fondamentale.

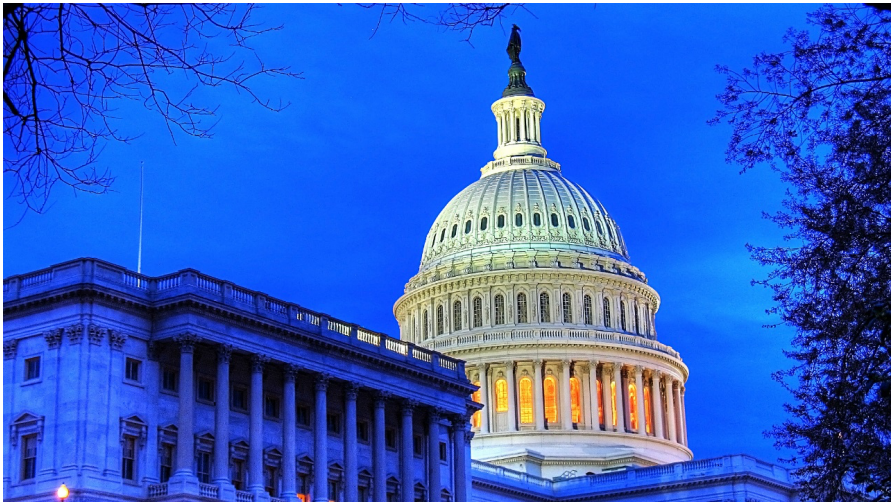
Le bâtiment du Capitole des États-Unis

Le deuxième domaine de tension concerne la méfiance croissante de la population à l'égard des scientifiques. Pour vous donner un exemple : Près de la moitié des Américains est profondément insatisfaite des politiques de lutte contre la pandémie de Covid, car elle était perçue comme une imposition par les scientifiques. Une sorte de charme a été rompu, et les scientifiques n'ont plus l'approbation universelle. Il n'est plus vrai que ce que les scientifiques disent qu'il faut faire suscite automatiquement le respect et résulte en d'excellentes perspectives de soutien financier.

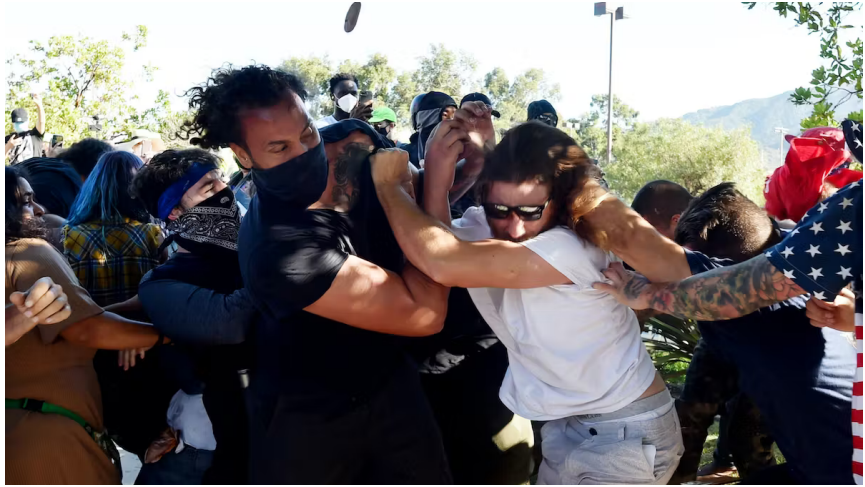

Le pays divisé (theconversation.com)

Le troisième domaine de tension est la crise budgétaire. Je suis assez âgé pour me souvenir de l'époque où un milliard de dollars représentait beaucoup d'argent. Ces dernières années, le Congrès a adopté des lois qui allouent des milliards de dollars aux infrastructures, définies de manière très large. Le ratio de la dette au PIB a augmenté rapidement. C'est extrêmement alarmant pour l'électorat américain, surtout compte tenu du lien que certains économistes dénoncent à la télévision entre ces dépenses supplémentaires, la dette croissante et l'inflation écrasante.

C'est donc une certitude que la majorité des candidats à la présidence américaine se présenteront sur une plate-forme électorale de réductions massives des dépenses discrétionnaires.

En d'autres termes, les projets scientifiques pourraient être supprimés, comme le président Trump voulait le faire et comme le président Bill Clinton l'a fait lorsqu'il a supprimé le Superconducting Supercollider en 1993 peu de temps après avoir pris ses fonctions sur une plate-forme électorale de réduction de la dette. Nous devrons être prêts à présenter des arguments solides en faveur de la valeur de la science fondamentale.

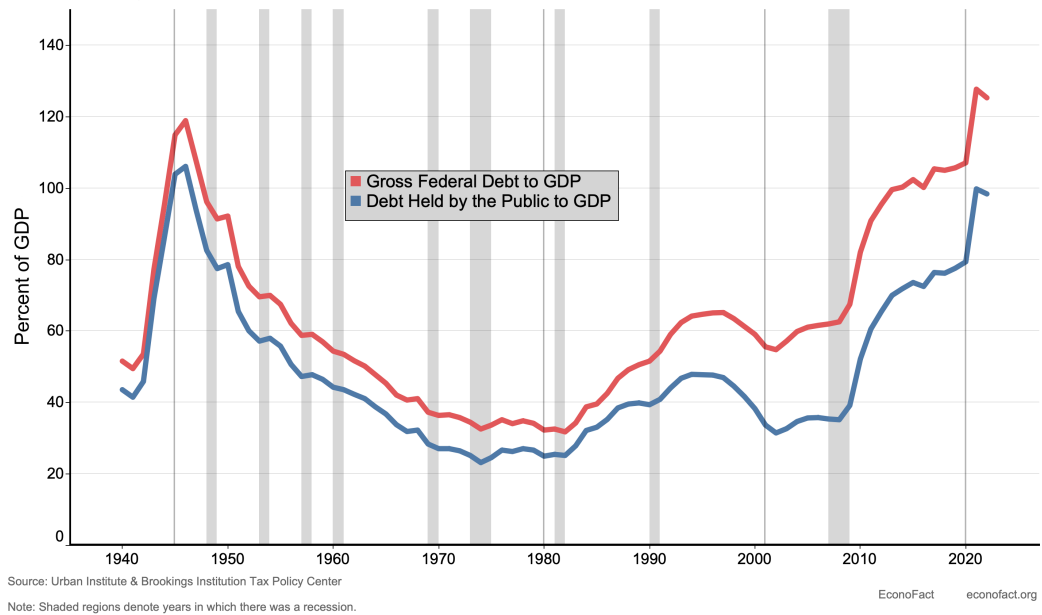

Econofact.org

*Participation pour tous*

Permettez-moi maintenant de passer à une brève discussion sur le changement politique qui est peut-être le plus important dans l'entreprise scientifique américaine de cette dernière décennie. Et c'est le sujet de la diversité, de l'équité et de l'inclusion, ou DEI comme il est souvent abrégé aux États-Unis. Cet effort était déjà en cours avant la reconnaissance officielle du président Biden par son décret exécutif de 2021 de « faire progresser la diversité, l'équité, l'inclusion et l'accessibilité au sein du gouvernement fédéral ». Peu de décrets ont été reçus avec autant de soutien passionné et de critiques enflammées dans un même temps.

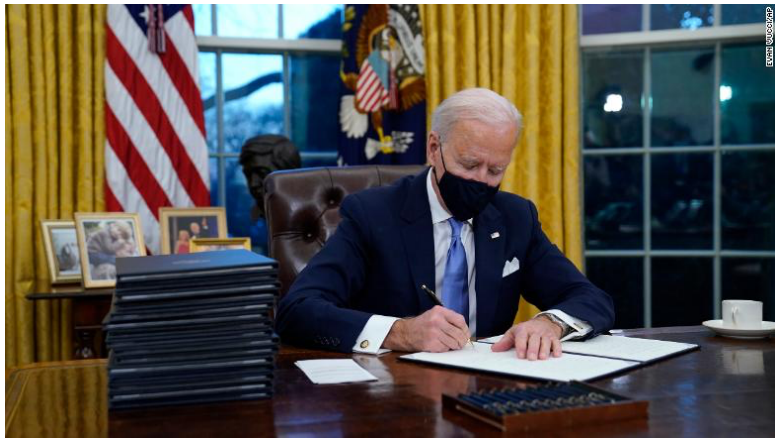

Président Joe Biden (cnn.com)

Un court paragraphe au début de l'annonce résume les objectifs du décret :
> « Même avec des décennies de progrès dans la construction d'une main-d'œuvre fédérale qui ressemble à l'Amérique, les héritages durables de la discrimination à l'emploi, du racisme systémique et de l'inégalité entre les sexes se font encore sentir aujourd'hui. Trop de communautés mal desservies restent sous-représentées dans la main-d'œuvre fédérale, en particulier dans les postes de direction. Ce décret établit une ambitieuse initiative pangouvernementale qui adoptera une approche systématique pour intégrer la DEIA …. »

Les effets se sont fait ressentir dans les universités, les laboratoires, les centres de recherche gouvernementaux, les sociétés scientifiques et dans l'industrie.

Des institutions scientifiques prestigieuses ont vu une augmentation importante de femmes et de minorité sous-représentées en leur seins et aux postes de direction ; et les sociétés scientifiques ont vu leur proportion de membres parmi ces groups augmenter à des taux jamais vu auparavant. Au nom du progrès « DEI », certains de ces processus ont contourné les critères plus traditionnels de sélection et de promotion. Celles-ci inclues les tests standardisés de connaissances et d'aptitudes, les étapes progressives d'accumulations d'expérience et les évaluations objectives d'impact scientifiques.

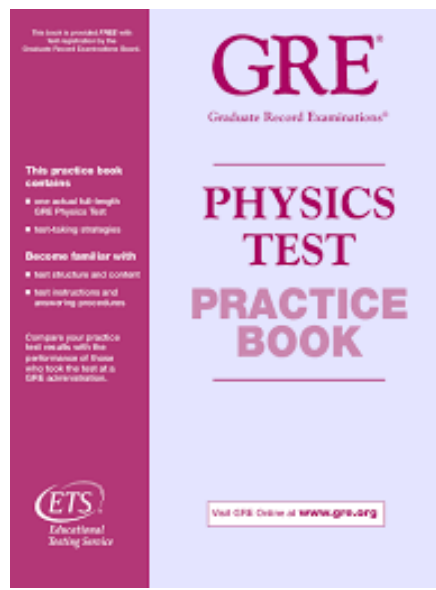

Examen d'entrée traditionnel
au doctorat de physique

Je ne souhaite pas exprimer une opinion tranchée sur ce sujet, car je pense qu'il est trop tôt pour dire quels seront les impacts finaux de cette importante poussée du DEI sur l'effort scientifique. Par contre, je souhaite énoncer quelques raisons pratiques pour lesquelles la DEI est nécessaire aux États-Unis, et peut-être ailleurs, également.

Certaines choses que je vais dire pourraient heurter certaines sensibilités françaises, car il y a une telle différence sur ces sujets entre la France et les États-Unis. C'est un anathème pour les Français de diviser et de catégoriser les gens aussi brutalement comme nous le faisons aux États-Unis. Comme l'a souligné Gérard Noiriel dans la préface de l'édition en langue anglaise de son célèbre livre « Le Creuset Français » :
> « Classer les gens, comme cela se fait en Californie, comme « blancs non hispaniques », « hispaniques », « noirs » et « asiatiques » serait considéré comme raciste en France, preuve d'une discrimination entre les citoyens français ; pour les Américains, cela semble naturel. »

Oui, en effet, c'est naturel pour nous. Dans le décret exécutif DEI du président Biden, il en impose davantage.

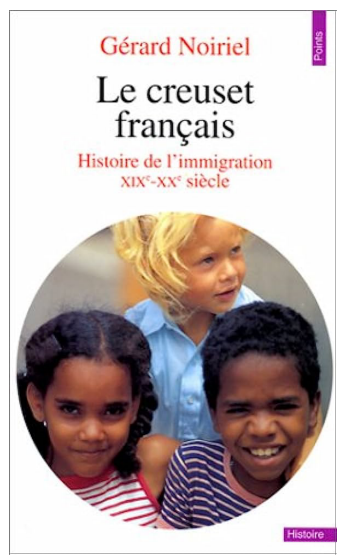

L'origine de ce système de classification est l'indignation que pour certaines personnes, la capacité de s'élever dans la société a été presque impossible en raison des identités et des circonstances dont elles ont hérité à la naissance. Lorsque nous voyons une telle injustice, nous ne sommes pas indignés parce que notre PIB national aurait pu connaitre une petite augmentation si ces personnes avaient eu une chance plus équitable de réussir. Non, nous nous indignons contre ce potentiel non réalisé. Comme le disait Antoine de Saint-Exupéry,

« Ce qui me tourmente … C'est un peu, dans chacun de ces hommes, Mozart assassiné. » et il faut ajouter ces femmes.

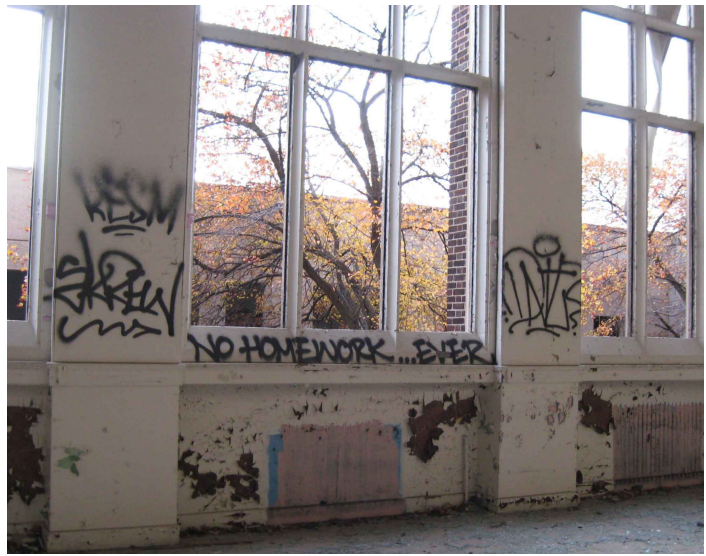
Une école à Detroit     (substancenews.net)

Sur le plan pratique, en outre, la constitution d'une coalition de soutien à la recherche scientifique de base nécessite une forte composante DEI aux États-Unis. Les finances publiques servent à soutenir les biens et services communs pour tous. Elles servent également à soutenir des initiatives sélectes. Pour beaucoup, les accélérateurs de particules et les télescopes spatiaux sont des exemples d'initiatives sélectes, comme les opéras chers et chics. De telles dépenses ne semblent profiter directement qu'à un petit nombre de personnes qui les réclament à cor et à cri. Ce qui est une dépense tolérable devient rapidement une dépense politiquement intolérable si des sous-groupes de personnes semblent être exclues de la participation à l'activité.

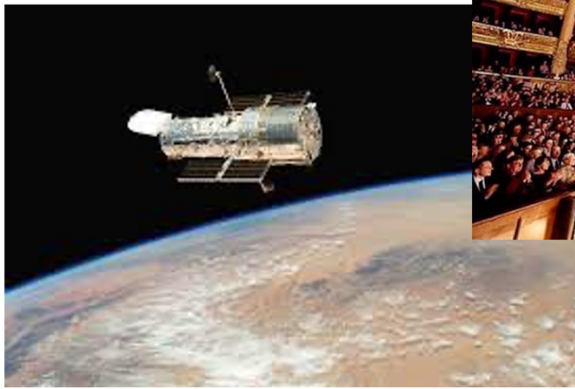
Hubble Space Telescope.     NASA

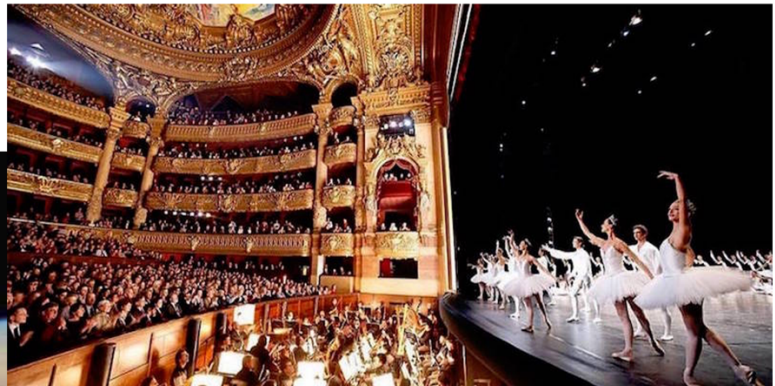
Palais Garnier

À tort ou à raison, la science fondamentale, comme d'autres initiatives sélectes, doit donc promouvoir la DEI non seulement parce que c'est la bonne chose à faire, mais aussi parce qu'elle est nécessaire à sa survie politique aux États-Unis.

Voilà, c'est ainsi que j'arrive à la fin de mon exposé. Je vous remercie de votre attention.